\documentclass[prb,aps,twocolumn,amsmath,amssymb,floatfix,pra,reprint,footinbib,superscriptaddress,showpacs,longbibliography]{revtex4-1}
\usepackage{graphics}      
\usepackage{graphicx}      
\usepackage{longtable}     
\usepackage{url}           
\usepackage{bm}    
\usepackage{braket}        
\usepackage{xcolor}
\usepackage{float}  
\usepackage{natbib}
\usepackage{graphicx}
\usepackage{comment}
\usepackage{siunitx}
\usepackage{appendix}
\usepackage{amsmath,amssymb,epsfig,color}
\usepackage{ulem}

\newcommand{\nn}{\nonumber}
\newcommand{\nl}{\nonumber \\}
\newcommand{\be}{\begin{equation}}
\newcommand{\ee}{\end{equation}}
\newcommand{\bea}{\begin{eqnarray}}
\newcommand{\eea}{\end{eqnarray}}

\begin{document}

\title {Classical Phase Space Crystals in an Open Environment}

\affiliation{Center for Joint Quantum Studies and Department of Physics, School of Science, Tianjin University, Tianjin 300072, China}
\affiliation{Szkoła Doktorska Nauk Ścisłych i Przyrodniczych, Wydział Fizyki, Astronomii i Informatyki Stosowanej, Uniwersytet Jagiello\'nski, ulica Profesora Stanisława Łojasiewicza 11, PL-30-348 Kraków, Poland}
\affiliation{Instytut Fizyki Teoretycznej, Wydział Fizyki, Astronomii i Informatyki Stosowanej, Uniwersytet Jagiello\'nski, ulica Profesora Stanisława Łojasiewicza 11, PL-30-348 Kraków, Poland}
\affiliation{Centrum Marka Kaca, Uniwersytet Jagiello\'nski, ulica Profesora Stanisława Łojasiewicza 11, PL-30-348 Kraków, Poland}
\affiliation{Max Planck Institute for the Science of Light, Staudtstrasse 2, 91058 Erlangen, Germany}

\author{Ali Emami Kopaei$^{2,3}$}
\author{Krzysztof Sacha$^{3,4}$}
\author{Lingzhen Guo$^{1,5}$}
\thanks{lingzhen.guo@mpl.mpg.de}

\begin{abstract}
It was recently discovered that a crystalline many-body state can exist in the phase space of a closed dynamical system. 
Phase space crystal can be anomalous Chern insulator that supports chiral topological transport without breaking physical time-reversal symmetry [L. Guo et al., Phys. Rev. B 105, 094301 (2022)].  
In this work, we further study the effects of open dissipative environment with  thermal noise, and identify the existence condition of classical phase space crystals in realistic scenarios. By defining a crystal order parameter, we plot the phase diagram in the parameter space of dissipation rate, interaction and temperature.
Our present work paves the way to realise phase space crystals and explore anomalous chiral transport in experiments. 
\end{abstract}

\date{\today}

\maketitle



\section{Introduction}
Physical systems in equilibrium are described by standard thermodynamics and statistical mechanics. For systems near equilibrium, linear response theory  \cite{Kubo1957JPSJ} applies by defining typical thermodynamic quantities locally, e.g., the Onsager reciprocal relations \cite{Onsager1931aps} and the principle of minimum entropy production \cite{Prigogine1945}.  However, physical systems far from equilibrium can behave drastically different. Nonequilibrium fluctuations can be amplified in the neighborhood of equilibrium stable point resulting in the so-called dissipative structures \cite{Prigogine1978science}, self-organisation phenomena \cite{Haken1983book} and chaotic structures, e.g., synchronisation \cite{Kuramoto1987jsp}, bifurcation \cite{May1976nature}, Lorenz attractor \cite{Lorentz1963JAS}, lasers,  Brusselator \cite{Prigogine1977book}, Rayleigh–B\'enard convection \cite{Getling1998book} and Belousov–Zhabotinsky reaction \cite{Hudson1982JCP}. These intriguing far-from-equilibrium phenomena have been studied intensively in classical dynamical systems for many decades and recently extended to the study in quantum systems such as quantum synchronisation \cite{Lee2013prl,Bruder2017prl,Weiss2017pra,Thomas2022pra} and period multiplication \cite{Delsing2017prb,Delsing2018apl,Arndt2022prb}. 
 
The novel far-from-nonequilibrium  states mentioned above are reached from the balance between driving (pumping energy) and damping (dissipating energy), i.e., by exchanging energy and information with an open environment. In contrast, the fate of a generic isolated driven many-body system is a trivial infinite temperature state \cite{Lazarides2014pre,DAlessio2014prx,Ponte2015AOP} due to the heating by the driving field. 
 One exceptional example is the \textit{Floquet/discrete time crystals} \cite{Sacha2018rpp,Khemani2019arxiv,Else2020arcmp,Guo2020njp} in a closed quantum system, where the \textit{discrete time transnational symmetry} (DTTS) of driving field is spontaneously broken and the infinite heating process is prevented by the disorder \cite{Khemani2016prl,Else2016prl,Yao2017prl} or effective nonlinearity in the thermodynamic limit \cite{Sacha2015pra,Russomanno2017prb,Giergiel2018pra,Matus2019pra,Giergiel2019prb}. For a clean system without disorder, there can also exist a prethermal state with an exponentially long lifetime if the driving frequency is much larger than the local energy scales \cite{Mori2016prl,KUWAHARA2016aop,Abanin2015prl,Abanin2015prl} resulting in the so-called \textit{prethermal time crystals}. By coupling the Floquet many-body system to a cold bath \cite{Kim2006prl,Heo2010pre}, the prethermal time crystal can have infinite lifetime and is dubbed as \textit{dissipative time crystals} \cite{Luitz2020prx,Else2017prx}. 
 While most studies focus on the spontaneous breaking of DTTS and the protection mechanism of time crystals, there is a trend to study the interplay of two or more time crystals \cite{Autti2021nm}, i.e., an emerging research field coined as \textit{condensed matter physics in time crystals} \cite{Sacha2018rpp,Guo2020njp,Hannaford2022aapps,Giergiel2018prl,Giergiel2020njp,Kuros2020njp,Giergiel2021prl,matus2021pra,Hannaford2022epl,Golletz2022njp,Kopaei2022pra}.
 
 Another example of ordered state in highly-excited system is the so-called \textit{phase space crystals} \cite{Guo2021book,Sacha2020book}
 , which is closely related to but different from time crystals. Depending on whether interaction is included, phase space crystals are classified as \textit{single-particle phase space crystals} and \textit{many-body phase space crystals} \cite{Guo2021book,Hannaford2022aapps}. 
 For a single-particle quantum system, the phase space crystal state refers to the eigenstate of Hamiltonian \cite{Guo2013prl} or the eigenoperator of the Liouvillian for an open quantum system \cite{Lang2021njp} that has discrete rotational or transnational symmetry in phase space. Phase space crystal in a many-body system is defined as the solid-like crystalline state in phase space \cite{Liang2018njp,Guo2022prb,Guo2021book}. In the work by  Guo et al. \cite{Guo2022prb}, the authors studied collective vibrational modes of many-body phase space crystals with a honeycomb lattice structure in phase space, and found the vibrational band structure can have nontrivial topological physics.
 Due to the symplectic phase-space dynamics,  the vibrations of any two atoms are coupled via a pairing interaction with intrinsically complex  phases that can not be eliminated by any local gauge transformation, leading to a vibrational band structure with non-trivial Chern numbers and chiral edge states in phase space. 
 In contrast to all the chiral transport scenarios in real space where the breaking of time reversal symmetry is a prerequisite, the chiral transport for phase space phonons can arise \textit{without} breaking physical time-reversal symmetry that becomes a global anti-unitary transformation in phase space.

In this work, we continue to investigate the classical dynamics of phase space crystals in open environment with dissipation and thermal noise. We  reduce the equation of motion (EOM) in the rest frame to the EOM with rotating wave approximation (RWA) in the rotating frame, which is then justified by numerical simulations. Based on the linear analysis of dynamical system, we find that the phase space crystals can exist when the interaction, dissipation and temperature are below some critical values.  We define an order parameter for phase space crystal state and plot the phase space diagram. Phase space crystals predicted by theory has not been found in the experiments. Our present work paves the way for the realisation of  phase space crystals in the ultra-cold atom experiment with realistic conditions.

The article is organized as follows. In Sec.~\ref{sec-ModelSystem}, we introduce the model system and the EOM in the open environment. In Sec.~\ref{sec-RotatingFrame}, we derive the EOM in the RWA  including dissipation and thermal noise.
In Sec.~\ref{sec-existence}, we study the dynamics of phase space crystals and identify the existence condition for the crystalline state in phase space. We first provide analytical results for the critical values of dissipation, temperature and interaction based on the linear analysis of dynamics. Then, we define the crystal order parameter and plot the phase diagram from numerical simulations based on RWA EOM. In Sec. V, we estimate the parameters for realizing classical phase space crystals in the real cold-atom experiments. In Sec.~\ref{sec-Summary}, we summarize the results in this work.

\begin{figure}
\center
\includegraphics[scale=0.45]{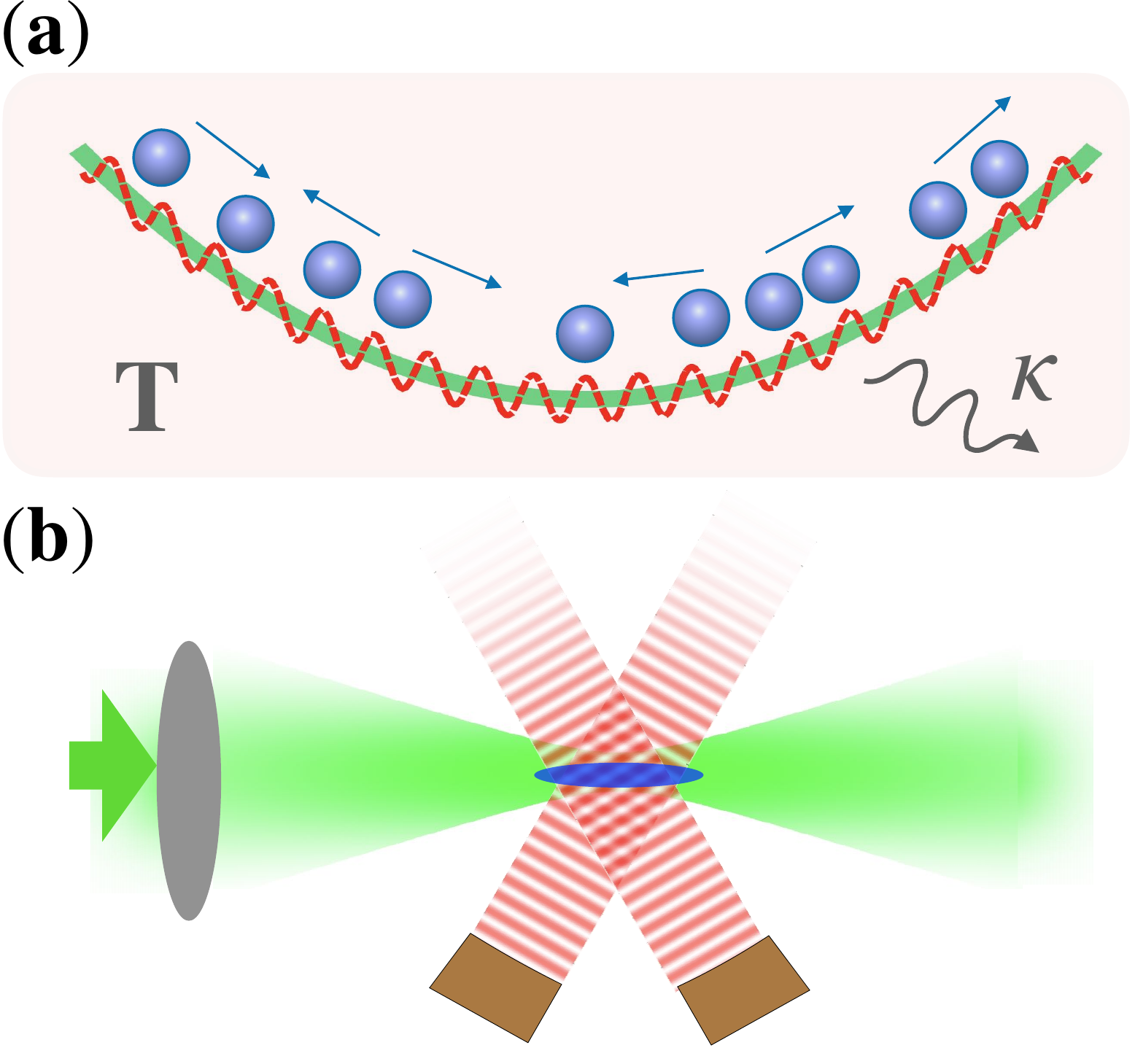}
\caption{{\bf  Model system.} {\bf  (a)} Sketch of our model: many particles (blue balls) moving in a harmonic trap (green curve) and periodically driven by additional lattice potential $V_d(x,t)$. The red dashed curve represents total potential at a fixed moment.  The whole system is subjected to a thermal bath with temperature $T$ and dissipation rate $\kappa$. {\bf (b)} Implementation of our model with cold atom experiment setup: a cloud of cold atoms (blue) confined in 1D harmonic trap formed by the intensity profile of a single Gaussian laser beam \cite{Bloch2008rmp} (green); the driven lattice potential is formed by two lasers (red) intersecting at angle with tunnable intensities and phases\cite{hadzibabic2004prl}. 
}\label{Fig-Modelsystem}
\end{figure}

\section{Model system}\label{sec-ModelSystem}

We consider the classical system of many particles trapped in a one-dimensional (1D) harmonic well and subjected to an additional periodically driven lattice potential $V_d(x,t)$ with driving frequency $\omega_d$ as shown in Fig.~\ref{Fig-Modelsystem}(a). In the experiment, such model system can be realised with cold atoms in optical lattices \cite{Moritz2003prl} as shown in Fig.~\ref{Fig-Modelsystem}(b).
At low temperature, the interaction of neutral cold atoms is dominated by $s$-wave scattering process and can be modelled by an effective two-body contact potential \cite{Bloch2008rmp}.
The total classical Hamiltonian of the system of $N$ atoms is given by
\bea\label{g}
H(t)&=&\sum_{i=1}^{N}\left[\frac{1}{2}(x_i^2+p_i^2)+V_d(x_i,t)\right]+\sum_{i<j}^N\beta\delta(x_i-x_j)\nl
&\equiv&\sum_{i=1}^NH_s(x_i,p_i)+\sum_{i<j}^N V(x_i,x_j),
\eea 
where $H_s(x_i,p_i)$ represents the single-atom Hamiltonian including the harmonic trap plus the driving potential, and $V(x_i,x_j)$ represents the real-space interaction of two atoms.
Here, all the variables have been scaled dimensionless by choosing the units of time, position and momentum as $\omega_0^{-1}$ ($\omega_0$ is the harmonic trapping frequency), $l_0$ (the characteristic length of driving lattice potential) and $p_0=m\omega_0l_0/2\pi$ ($m$ is the mass of particle) respectively. The unit of energy (Hamiltonian) is set to be $\epsilon_0=m\omega_0^2(l_0/2\pi)^2$. 

In the open environment with dissipation rate $\kappa$ (scaled by $\omega_0$) and temperature $T$ (scaled by $\epsilon_0/k_B$ with $k_B$ the Boltzmann constant), the classical EOM is given by
\begin{equation}\label{EEOM}
\frac{dx_i}{dt}= p_i,\ \ \
\frac{dp_i}{dt}=-\frac{\partial }{\partial x_i}H(t)-\kappa p_i+\sqrt{2\kappa T}n_i(t).
\end{equation}
Here, the thermal noise term $\sqrt{2\kappa T}n_i(t)$ is introduced according to the fluctuation-dissipation relationship, where $n_i(t)$ is the white noise satisfying 
\begin{eqnarray}\label{eq-xit}
\overline{n_i(t)}=0,\ \ \ \overline{n_i(t)n_j(t')}=\delta_{ij}\delta(t-t').
\end{eqnarray}
 We now define the Wiener process as the integral of white noise, i.e.,
 $w_i(t)\equiv\int_0^tn_i(\tau)d\tau.$
Using the property of white noise Eq.~(\ref{eq-xit}), one can show that $$\overline{w_i(t)}=0,\ \  \overline{w_i(t)w_i(t')}=\delta_{ij}\mathrm{min}(t,t').$$ By further defining $dw_i(t)\equiv w_i(t+dt)-w_i(t)$, we have
$$\overline{dw_i(t)}=0,\ \ \ \overline{dw_i^2(t)}=\overline{[w_i(t+dt)-w_i(t)]^2}=dt.$$
Therefore, we can write the EOM given by Eq.~(\ref{EEOM}) as the following stochastic differential equation process
\begin{eqnarray}\label{eq-stxp}
\left\{\begin{array}{l}
dx_i(t)= p_i(t)dt, 
\\
dp_i(t)=-\frac{\partial H}{\partial x_i}dt-\kappa p_i(t)dt+\sqrt{2\kappa T}dw_i(t).
\end{array}\right.
\end{eqnarray}

\section{Rotating frame}\label{sec-RotatingFrame}

\subsection{ Hamiltonian within rotating wave approximation}

We go to the rotating frame with frequency $\Omega/n$ using the generating function of the second kind
\begin{equation}\label{G2}
G_2(t)=\sum_i\frac{x_iP_i}{\cos(\Omega t/n)}-\frac{1}{2}x_i^2\tan\Big(\frac{\Omega}{n}t\Big)-\frac{1}{2}P_i^2\tan\Big(\frac{\Omega}{n}t\Big).\nn
\end{equation}
Here, we have defined the ratio of driving frequency to harmonic frequency by $\Omega=\omega_d/\omega_0$ and assumed the near-resonance condition $\Omega\sim n$ with $n\in \mathbb{Z}^{+}$.
Note that we can introduce
some detuning $\delta\omega=1-\Omega/n$ between the driving and harmonic frequencies if $\Omega \neq n$, which will produce a parabolic confinement potential in phase space that is sometimes important to stabilise the phase space crystals \cite{Guo2022prb}.  
The canonical transformation of coordinates is then given by 
$p_i=\partial G_2/\partial x_i$, $X_i=\partial G_2/\partial P_i$, i.e.,
\begin{eqnarray}\label{eq-xipi}
\left\{\begin{array}{l}
x_i(t)=P_i\sin\big(\Omega t/n\big)+X_i\cos\big(\Omega t/n\big), 
\\
p_i(t)=P_i\cos\big(\Omega t/n\big)-X_i\sin\big(\Omega t/n\big).
\end{array}\right.
\end{eqnarray}
The canonical transformation of Hamiltonian Eq.~(\ref{g}) in the rotating frame is given by 
\bea
H_{RF}(t)=H(t)+\partial G_2(t)/\partial t.
\eea
Due to the driving field and interaction of atoms, the quadratures $(X_i,P_i)$ of oscillation (amplitude and phase) are slowly moving. By plugging the transformation Eq.~(\ref{eq-xipi}) into $H_{RF}(t)$ and neglecting all the time-dependent (fast oscillating) terms, we arrive at the effective static Hamiltonian in the rotating wave approximation (RWA) 
\bea
\label{eq-HRAW}
\mathcal{H}=\sum_i\mathcal{H}_s(X_i,P_i)+\sum_{i<j}U(R_{ij}).
\eea
We expect the RWA is valid when the driving field $V_d(x_i,t)$ and the contact interaction strength $\beta$ between atoms are weak compared to the harmonic trapping frequency, which will be justified by our numerical simulation in Sec.~\ref{sec:justtification}.
Here, $\mathcal{H}_s(X_i,P_i)$ represents the RWA part of single-atom Hamiltonian $H_s(x_i,p_i)$, cf., Eq.~(\ref{g}). 

For short-range interactions in real space, the effective RWA interaction becomes a function  of the distance between atoms in phase space \cite{Guo2016pra,Guo2020njp} 
\bea\label{eq-Rij}
R_{ij}=\sqrt{(X_i-X_j)^2+(P_i-P_j)^2}.
\eea
This is because the atoms located at different phase space points will still collide in the course of their laboratory-frame trajectories. Thus, when we perform averaging of the Hamiltonian over time, the short-range interaction in the laboratory-frame gives rise to an effective long-range interaction in the rotating frame \cite{Sacha2015sr,Sacha2015pra,
Guo2016pra,Giergiel2018prl,Liang2018njp,Guo2021book}.
For the contact interaction of cold atoms in the laboratory frame, the effective interaction becomes long-range Coulomb-like interaction\cite{Guo2016pra,Liang2018njp,Guo2021book}
\bea\label{eq-URij}
U(R_{ij})=\frac{\beta}{\pi}\frac{1}{R_{ij}}.
\eea

\begin{figure*}
\center
\includegraphics[scale=0.7]{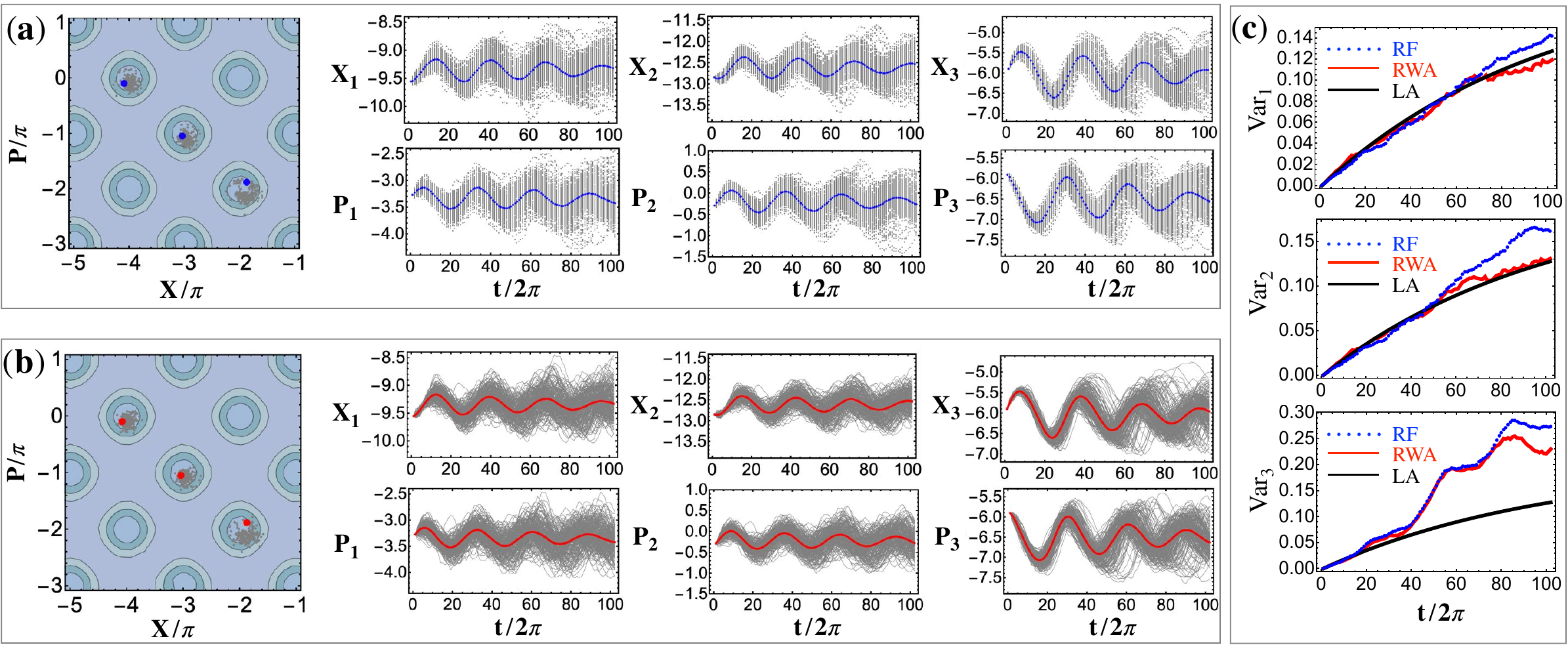}
\caption{{\bf Justification of RWA in the presence of dissipation and thermal noise.} {\bf(a)} Stroboscopic dynamics of  $N=3$ interacting particles in the laboratory frame obtained by solving  exact EOM (\ref{eq-stxp}) and plotting the phase space variables every period of the harmonic oscillator potential. The three blue dots in the leftmost panel indicate the initial conditions of three particles. The other panels show 200 trajectories of each particle (grey dots) obtained for 200 different realizations of noise and the averaged dynamics (blue dots).  {\bf(b)} Dynamics of three interacting particles in the rotating frame obtained within RWA by solving Eqs.~(\ref{SEOMwithDF}). The three red dots in the leftmost panel indicate the initial conditions of three particles. The other panels show 200 trajectories of each particle (grey trajectories) obtained for 200 different realizations of noise and the averaged dynamics (red curves).   {\bf(c)} Time evolution of the variance of each particle in phase space, cf. Eq.~(\ref{eq-vari}), obtained from the exact EOM (blue dots) and within RWA (red curve) and linear analysis (LA) (black curve), cf. Eq.~(\ref{Var_i_analytical}).
For all figures, we set driving strength $\Lambda=-0.01$, dissipation rate $\kappa/\kappa_c=0.5$, interaction $\beta/\beta_c=0.5$ with the Lorenz parameter $\epsilon=0.5$, cf. Eq.~(\ref{eq-URepsilon}), and temperature $T=0.1$.
}\label{Fig-EOM}
\end{figure*}

\subsection{Equations of motion within rotating wave approximation}

For a closed system without dissipation, the canonical EOM under RWA in the rotating frame is given by
\begin{eqnarray}\label{EOM}
\frac{dX_i}{dt}=\frac{\partial \mathcal{H}}{\partial P_i}, \ \ \ \ \ \
\frac{dP_i}{dt}=-\frac{\partial \mathcal{H}}{\partial X_i} .
\end{eqnarray}
In order to obtain the EOM under RWA in the open environment with dissipation and thermal noise, we introduce the following transformation from Eq.~(\ref{eq-xipi})
\begin{eqnarray}\label{eq-dXidPi}
\left\{\begin{array}{l}
\frac{dX_i}{dt}=\frac{dx_i}{dt}\cos\Big(\frac{\Omega}{n}t\Big)-\frac{dp_i}{dt}\sin\Big(\frac{\Omega}{n}t\Big)-\frac{\Omega}{n}P_i
\\
\frac{dP_i}{dt}=\frac{dx_i}{dt}\sin\Big(\frac{\Omega}{n}t\Big)+\frac{dp_i}{dt}\cos\Big(\frac{\Omega}{n}t\Big)+\frac{\Omega}{n}X_i.
\end{array}\right.
\end{eqnarray}
By plugging Eq.~(\ref{EEOM}) into Eq.~(\ref{eq-dXidPi}), we have EOM including the dissipation and noise in the rotating frame
\begin{eqnarray}\label{eq-nXnP}
\left\{\begin{array}{l}
\frac{dX_i}{dt}=\begin{array}{c}\frac{\partial H_{RF}}{\partial P_i}\end{array}+\kappa p_i\sin\Big(\frac{\Omega}{n}t\Big)-\sqrt{\kappa T}n^X_i(t)
\\
\frac{dP_i}{dt}=\begin{array}{c}-\frac{\partial H_{RF}}{\partial X_i}\end{array}-\kappa p_i\cos\Big(\frac{\Omega}{n}t\Big)+\sqrt{\kappa T}n^P_i(t).
\end{array}\right.
\end{eqnarray}
Here, we have defined the two orthogonal components of noises by
\bea
n^X_i(t)\equiv \sqrt{2}n_i(t)\sin\Big(\frac{\Omega}{n}t\Big), \;\; n^P_i(t)\equiv \sqrt{2}n_i(t)\cos\Big(\frac{\Omega}{n}t\Big).\nn
\eea
 Obviously, we have $\overline{n^X_i(t)}=0$ and $\overline{n^P_i(t)}=0$. From the relationship Eq.~(\ref{eq-xit}), we have the time correlation of two noise components as follows
\begin{eqnarray}\label{FDT3}
&&\overline{n^X_i(t)n^P_j(t')}
=\delta_{ij}\delta(t-t')\sin\Big(2\frac{\Omega}{n}t\Big)\\
&&  \overline{n^X_i(t)n^X_j(t')}=\delta_{ij}\delta(t-t')\Big[1-\cos\Big(2\frac{\Omega}{n}t\Big)\Big]\\
&&   \overline{n^P_i(t)n^P_j(t')}=\delta_{ij}\delta(t-t')\Big[1+\cos\Big(2\frac{\Omega}{n}t\Big)\Big].
\end{eqnarray}
Therefore, in the RWA (keeping only time-independent terms in the correlations), we can take $n^X_i(t)$ and $n^P_i(t)$ as independent standard white noises. 

We plug Eq.~(\ref{eq-xipi}) into EOM (\ref{eq-nXnP}) and keep only static term in the spirit of RWA. Finally, we obtain the RWA EOM with dissipation and noise 
\begin{eqnarray}\label{EOMwithDF}
\left\{\begin{array}{c}\frac{dX_i}{dt}=\frac{\partial \mathcal{H}}{\partial P_i}- \frac{1}{2}\kappa X_i-\sqrt{\kappa  T}n^{X}_i(t) \\
\ \ \  \frac{dP_i}{dt}=-\frac{\partial \mathcal{H}}{\partial X_i}- \frac{1}{2}\kappa P_i+\sqrt{\kappa  T}n^{P}_i(t).
\end{array}\right.
\end{eqnarray}
As in Eq.~(\ref{eq-stxp}), we introduce two independent Wiener processes for the two quadratures $w^{X}_i(t)=\int_0^tn^X_i(\tau)d\tau$ and $w^{P}_i(t)=\int_0^tn^P_i(\tau)d\tau$ and write the EOM (\ref{EOMwithDF}) within RWA in the form of stochastic differential equation 
\begin{eqnarray}\label{SEOMwithDF}
\left\{\begin{array}{c}
dX_i(t)=\big(\begin{array}{c}\frac{\partial \eta}{\partial P_i}\end{array}- \frac{1}{2}\kappa X_i\big)dt-\sqrt{\kappa T}dw^{X}_i(t), \\
\ \  dP_i(t)=\big(\begin{array}{c}-\frac{\partial \eta}{\partial X_i}\end{array}- \frac{1}{2}\kappa P_i\big)dt+\sqrt{\kappa T}dw^{P}_i(t).
\end{array}\right.
\end{eqnarray}

\subsection{Justification}\label{sec:justtification}

In order to justify our RWA  with dissipation and thermal noise, we consider the following classical many-body Hamiltonian with square phase space lattice described by
\begin{eqnarray}\label{eq-square}
\mathcal{H}=\sum_{i=1}^N\Lambda(\cos X_i+\cos P_i)^2+\sum_{i<j}^NU(R_{ij}).
\end{eqnarray}
The square lattice of single-particle  Hamiltonian in phase space can be generated by a kicking sequence of stroboscopic lattices (see details in Appendix)
\bea
V_d(x,t)=\Lambda\sum_{n\in\mathbb{Z}}\sum_qK_q\cos(k_qx)\delta\left(\frac{t}{2\pi}-\theta_q-n\right).
\eea
%
Here, there are six kicks in each harmonic time period with kicking parameters $k_q =[ \sqrt{2},2,-\sqrt{2},-\sqrt{2},\sqrt{2},2 ]$, $\theta_q =[1/8,2/8,3/8,5/8,7/8,1]$ and $K_q=0.5\Lambda$.
In fact, arbitrary lattice structures in phase space can be synthesized by properly choosing the kicking parameters \cite{Guo2022prb,Guo2021book}.

The validity of RWA EOM without dissipation has been studied in the previous works \cite{Guo2016pra,Liang2018njp}. Here, we justify our RWA EOM with finite dissipation rate and at finite temperature by comparing the prediction of Eqs.~(\ref{SEOMwithDF}) with numerical solutions of the exact EOM (\ref{EEOM}). In our numerical simulation, we choose Lorenz function to model the contact interaction potential, i.e.,
 $V(x_i,x_j)=\lim_{\epsilon\rightarrow 0}\frac{\beta}{\pi}\frac{\epsilon}{(x_i-x_j)^2+\epsilon^2}$. The corresponding phase space interaction potential is then given by \cite{Guo2016pra,Liang2018njp}
 \bea\label{eq-URepsilon}
 U(R_{ij})=\lim_{\epsilon\rightarrow 0}\frac{\beta}{\pi}\frac{1}{\sqrt{R_{ij}^2+\epsilon^2}}.
 \eea
In Fig.~\ref{Fig-EOM}(a)-(b), we compare the dynamics  of $N=3$ interacting particles obtained within RWA with the exact results obtained by solving Eqs.~(\ref{eq-stxp}). We show 200 trajectories corresponding to 200 realizations of noise for the same initial conditions of the particles, and also present the averaged phase space variables $(\overline{X}_i(t),\overline{P}(t)_i)$. Clearly, the $200$ samples spread gradually as time evoluates. In Fig.~\ref{Fig-EOM}(c), we plot the variance of threes particles in the phase space given by (cf. also Eqs.~(\ref{varu})-(\ref{Var_i_analytical}))
\bea\label{eq-vari}
\mathrm{Var}_i(t)\equiv\overline{( { X}_i-{ \overline{X}}(t)_i)^2}+\overline{( { P}_i-{ \overline{P}}(t)_i)^2}.
\eea
The numerical results show that the RWA approach agrees well with the exact dynamics.  In fact, the RWA is valid when the dynamics of $(X_i,P_i)$ is much slower than the period of the harmonic oscillator potential.

\section{Existence of phase space crystals}\label{sec-existence}

In the present section, we analyse the dynamics of phase space crystals based on the many-body dynamics and in the presence of dissipation and thermal noise and within the RWA, Eq.~(\ref{EOMwithDF}).  Our goal is to identify the condition for the existence of phase space crystals.
According to Eq.~(\ref{eq-Rij}), the RWA interaction potential of two atoms  $U(R_{ij})$ depends on their relative distance in the phase space. Therefore,
it is natural to extend the concept of force from configuration space to phase space. By defining the position vector in phase space ${\bf Z}_i\equiv (X_i,P_i)^T$ and a unit direction vector perpendicular to the phase space plane ${\bf \hat{n}}\equiv{\bf \hat{n}}_X\times{\bf \hat{n}}_P$, where ${\bf\hat{n}}_X$ and ${\bf \hat{n}}_P$ are the unit vectors in the position and momentum directions respectively, we can rewrite the EOM~(\ref{EOMwithDF}) in the following compact form
\bea\label{ZEOMNoise}
\frac{d}{dt}{\bf Z}_i={\bf \hat{n}}\times {\bf F}_i,
\eea
with the \textit{phase space force} defined via
\bea\label{eq-FikT}
{\bf F}_i\equiv -{\bf \nabla}_i\mathcal{H} +\frac{1}{2}\kappa {\bf \hat{n}}\times {\bf Z}_i+\sqrt{\kappa T}{\bf n}_i(t).
\end{eqnarray}
Here, we have defined the thermal noise vector $${\bf n}_i(t)\equiv\big(n^{X}_i(t),n^{P}_i(t)\big)^T.$$ Based on the linear analysis of Eqs.~(\ref{ZEOMNoise}) and (\ref{eq-FikT}), we will estimate the critical values of relevant parameters (dissipation rate, temperature and interaction strength) for the existence of phase space crystals, which will be  verified by numerical simulations.


\subsection{Dissipation effects}

For a closed system of particles without interaction, the fixed points ${\bf Z}^0_i$ are the extreme points determined by the condition  ${\bf \nabla}_i\mathcal{H} |_{ {\bf Z}_i={\bf Z}^0_i}=0.$ We have  ${\bf Z}^0_i=(n\pi,m\pi)^T$ with  $n,m\in\mathbb{Z}$ having the same parity.
Finite values of dissipation rate, temperature and the presence of interaction will shift the fixed points and thus affect the existence of phase space crystals.
We first consider the pure effects of dissipation at zero temperature ($T=0$)
and without interaction ($\beta=0$). In this case, the fixed points in phase space are given by the condition of $ {\bf F}_i=0$. From Eq.~(\ref{eq-FikT}), we have
\begin{eqnarray}\label{cofnfp}
{\bf \nabla}_i\mathcal{H} \big|_{{\bf Z}_i={\bf \tilde{Z}}^0_i}= \frac{1}{2}\kappa \big({\bf \hat{n}}\times{\bf \tilde{Z}}^0_i\big),
\end{eqnarray}
where $ {\bf \tilde{Z}}^0_i$ are the positions of the fixed points shifted by dissipation. 
We linearize Eq.~(\ref{cofnfp}) around the original fixed points ${\bf Z}^0_i$ as follows 
\begin{eqnarray}\label{linear1}
{\bf J}( {\bf \tilde{Z}}^0_i-{\bf Z}^0_i)=\frac{1}{2}\kappa {\bf A} ({\bf \tilde{Z}}^0_i-{\bf Z}^0_i)+\frac{1}{2}\kappa {\bf A} {\bf Z}^0_i,
\end{eqnarray}
where the Jacobian matrix ${\bf J}$ and asymmetric tensor ${\bf A}$ are given by 
\begin{eqnarray}\label{JandA}
{\bf J}&\equiv&{\bf J}_{\nabla_i\mathcal{H}}|_{ {\bf Z}_i={\bf Z}^0_i}\equiv\left(\begin{array}{cc}\frac{\partial^2\mathcal{H}}{\partial X_i\partial X_i} & \frac{\partial^2\mathcal{H}}{\partial X_i\partial P_i} 
\\\frac{\partial^2\mathcal{H}}{\partial P_i\partial X_i} & \frac{\partial^2\mathcal{H}}{\partial P_i\partial P_i}\end{array}\right)\Bigg\vert_{ {\bf Z}_i={\bf Z}^0_i}, \label{eq-JDeta}\\
{\bf A}&\equiv&\left(\begin{array}{cc}0 & -1 
\\1 & 0\end{array}\right).
\end{eqnarray}
Solving Eq.~(\ref{linear1}), we get the shifted fixed points explicitly
\begin{eqnarray}\label{ZiZ02}
{\bf \tilde{Z}}^0_i-{\bf Z}^0_i=- \frac{2\Lambda \kappa}{16\Lambda^2+\kappa^2/4}{\bf \hat{n}} \times{\bf Z}^0_i-\frac{\kappa^2/4}{16\Lambda^2+\kappa^2/4}{\bf Z}^0_i.\nl
\end{eqnarray}
\noindent  
The first term on the right hand side of Eq.~(\ref{ZiZ02})
is responsible for rotation of the whole lattice while the second term contracts the whole lattice along the radial direction.

\begin{figure}
\center
\includegraphics[scale=0.35]{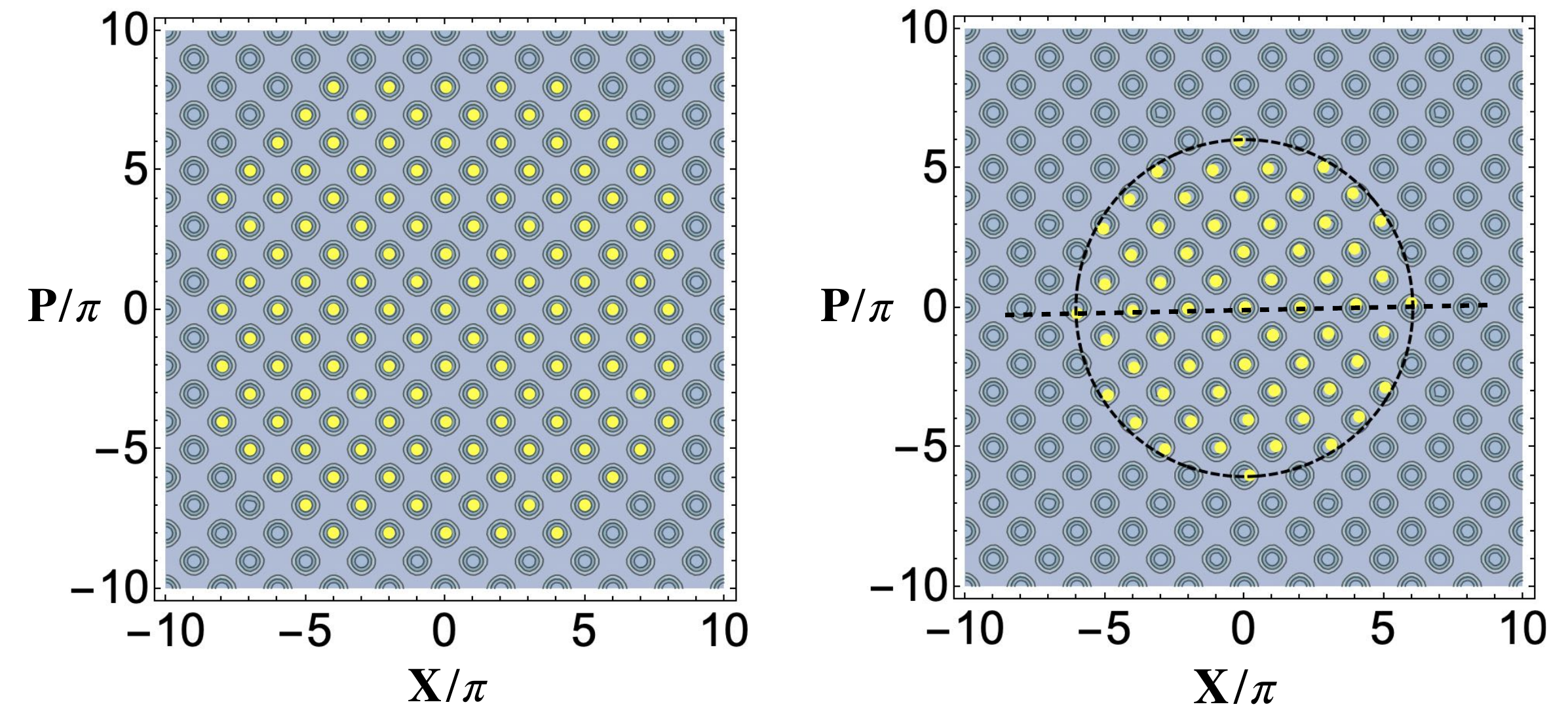}
\caption{{\bf Dissipation effects on phase space crystal formation.} {(Left):} Initial conditions of $N=129$ atoms (yellow dots). {(Right):} Final configuration of atoms with dissipation rate $\kappa=1.5\kappa_c$. The dashed circle and dashed line indicate the crystal radius predicted by Eq.~(\ref{eq-Rc}) and the rotation angle given by Eq.~(\ref{ZiZ02}). Driving strength parameter: $\Lambda=-0.01$.
}\label{Fig-Dissipation}
\end{figure}

If the shift of the fixed points is large enough, one can imagine that atoms will escape the lattice potential and the phase space crystals become ``melting". Since the displacement of the fixed points is proportional to their distance from the origin, cf. Eq.~(\ref{ZiZ02}), the crystal will start to melt from the edge. In order to calculate the critical dissipation rate where the atoms on the edge start to melt, we estimate the size of the final lattice limited due to dissipation. From Eq.~(\ref{cofnfp}), we have the following condition
\begin{eqnarray}\label{exiscond1}
 &&\Big|\frac{1}{2}\kappa \big({\bf \hat{n}}\times{\bf Z_i}\big)\Big|
 =|{\bf \nabla}_i\mathcal{H}|\nl
 &=&2 \Big|\Lambda(\cos X_i+\cos P_i)\Big|\sqrt{\sin^2 X_i+\sin^2 P_i}\nl
 &\leqslant& 2\sqrt{2}\big|\Lambda\big|.
\end{eqnarray}
The above equation sets an upper limit for the lattice size condition, i.e. solutions for the fixed points exist if
$ |{\bf Z_i}|\leq R_c$ where $ R_c\sim 4\sqrt{2}{\Lambda}/{\kappa}$.
Considering the angular dependence of $|{\bf \nabla}_i\mathcal{H}|$, we get a better empirical estimation for the radius of stable region  from numerical simulations, 
\bea\label{eq-Rc}
R_c\sim \frac{4\Lambda}{\kappa}.
\eea
The area of stable region for the existence of phase space crystals is  approximately $$S=\pi R^2_c=\pi({4\Lambda}/{\kappa})^2.$$ 
The total number of  the fixed points inside the stable region is approximately $$\sigma_a S=\frac{8}{\pi}\Big(\frac{\Lambda}{\kappa}\Big)^2,$$ where $\sigma_a={2}/{(2\pi)^2}$ is the density of atoms, e.g., one atom in each fixed point which corresponds to two atoms in each unit cell.
If we assume that each fixed point is occupied by one atom, there is an upper limit for total atom number 
$
 N=\frac{8}{\pi}\Big(\frac{\Lambda}{\kappa}\Big)^2.
$ 
Given the number of atoms, the critical dissipation rate is 
\bea
\kappa_c=\frac{4|\Lambda|}{\sqrt{2\pi N}}.
\eea
In Fig.~\ref{Fig-Dissipation}, we show the initial state of atoms (left) and the final state due to the dissipation (right). The black dashed circle indicates the crystal radius predicted by Eq.~(\ref{eq-Rc}).
The dashed line indicates the  rotation angle of the final lattice given by our predication, c.f. Eq~(\ref{ZiZ02}). 
Note that the two plots have the same number of atoms. As we do not consider interaction here, it is possible that more than one atoms occupy the same lattice site in the final crystal state.

\subsection{Temperature effects}

We then study the effects of thermal noise that is determined by a finite temperature $T>0$ and dissipation rate $\kappa>0$, cf. Eq.~(\ref{eq-FikT}). We apply linear approximation around the points ${\bf Z}^0_i$ to the EOM (\ref{ZEOMNoise}) 
\begin{eqnarray}\label{ZEOMNoise2}
\frac{d}{dt}({\bf Z}_i-{\bf Z}^0_i)&=& -\big({\bf AJ+\frac{1}{2}\kappa I}\big)({\bf Z}_i-{\bf Z}^0_i)\nl
&&-\frac{1}{2}\kappa  {\bf Z}^0_i+\sqrt{\kappa T}{\bf A}n_i(t).
\end{eqnarray}
%
 We now define the following auxiliary vector 
\begin{eqnarray}\label{Au}
{{\bf u}}_i\equiv{\bf A}({\bf Z}_i-{\bf \tilde{Z}}^0_i),
\end{eqnarray}
where ${\bf \tilde{Z}}^0_i$ is the shifted equilibrium point due to dissipation but at $T=0$, cf., Eq.~(\ref{linear1}) or (\ref{ZiZ02}). The vector ${{\bf u}}_i$ is perpendicular to the displacement vector describing the shift ${\bf Z}_i-{\bf \tilde{Z}}^0_i$. We simplify the stochastic differential equation (\ref{ZEOMNoise2}) as follows
\begin{eqnarray}\label{}
d{\bf { u}}_i(t)
=-{\bf B}{\bf {u}}_i(t)dt-\sqrt{\kappa T}d{\bf w}_i(t),
\end{eqnarray}
with
\begin{eqnarray}\label{eq-bfB}
{\bf B}\equiv\big({{\bf J}{\bf A}+\frac{1}{2}\kappa {\bf I}}\big),\ \ 
d{\bf w}_i(t)={\bf n}_i(t)dt.
\end{eqnarray}
This stochastic differential equation describes a multi-dimensional Ornstein-Uhlenbeck process. 
The formal solution is given by
\begin{eqnarray}\label{}
{\bf { u}}_i(t)
=e^{{-\bf B}t}{\bf {u}}_i(0)-\sqrt{\kappa T}\int_0^{t}e^{{-\bf B}(t-s)}d{\bf w}_i(s).
\end{eqnarray}
The mean value of stochastic process ${\bf { u}}_i(t)$ is 
\begin{eqnarray}\label{}
\mathrm{E}[{\bf{ u}}_i(t)]
=e^{{-\bf B}t}{\bf {u}}_i(0)={\bf P}{\bf E}(t){\bf P}^{-1}{\bf {u}}_i(0)
\eea
with
\bea
{\bf E}(t)\equiv\left(\begin{array}{cc}e^{-\lambda_1t} & 0 \\0 & e^{-\lambda_2t}\end{array}\right).
\end{eqnarray}
Here, {\bf P} is the matrix diagonalising the matrix ${\bf B}$, i.e., ${\bf P}^{-1}{\bf B}{\bf P}=\mathrm{diag}(\lambda_1,\lambda_2)$. 

For open dissipative  environment, the eigenvalues $\lambda_{1}$ and $\lambda_2$ should be positive values such that $\mathrm{E}[{\bf{ u}}_i(t)]\rightarrow 0$ in the long time limit. We further calculate the variance of  ${\bf{ u}}_i(t)$ as follows
\begin{eqnarray}\label{varu}
\mathrm{Var}_i
&=&\mathrm{E}\big[\big({\bf{ u}}_i(t)-\overline{{\bf{ u}}_i(t)}\big)^T\big({\bf { u}}_i(t)-\overline{{\bf{ u}}_i(t)}\big)\big]\nonumber\\
&=&\kappa T\int_0^{t}\int_0^{t}\mathrm{E}\big[d{\bf w}^T_i(s')e^{{-\bf B}^T(t-s')}e^{{-\bf B}(t-s)}d{\bf w}_i(s)\big]
\nonumber\\
&=&\kappa T\int_0^{t}ds\mathrm{E}\big[{\bf n}^T_i(s)e^{{-\bf B}^T(t-s)}e^{{-\bf B}(t-s)}{\bf n}_i(s)\big].
\end{eqnarray}
Here, we have used the property of white noise Eq.~(\ref{eq-xit}). 
Reminiscent of the  definition of ${\bf B}$ given by Eq.~(\ref{eq-bfB}), it is not difficult to prove the following statement:
\textit{If matrix ${\bf J}$ is diagonal, we have the identities: ${\bf JA}={\bf AJ}$, $[{\bf B},{\bf B}^T]=0.$}
 The Jacobian matrix, Eq.~(\ref{JandA}), is indeed diagonal, ${\bf J}={\rm diag}(-4\Lambda,-4\Lambda)$, and thus $[{\bf B},{\bf B}^T]=0$ 
 holds in our case. Therefore, according to the Baker-Campbell-Hausdorff formula, we have 
$$e^{{-\bf B}^T(t-s)}e^{{-\bf B}(t-s)}=e^{-({\bf B}^T+{\bf B})(t-s)}=e^{-\kappa(t-s)}{\bf I}.$$
As a result, the variance of stochastic process ${\bf{ u}}_i(t)$ given by Eq.~(\ref{varu}) can be calculated explicitly as follows
\begin{eqnarray}\label{Var_i_analytical}
\mathrm{Var}_i
&=&\mathrm{E}\big[\big({\bf{ u}}_i(t)-\overline{{\bf{ u}}_i(t)}\big)^T\big({\bf { u}}_i(t)-\overline{{\bf{ u}}_i(t)}\big)\big]\nl
&=&2\kappa T\int_0^{t}dse^{-\kappa(t-s)}\nl
&=&2T (1-e^{-\kappa t}).
\end{eqnarray}
In Fig.~\ref{Fig-EOM}(c), we compare the variance (\ref{Var_i_analytical}) with the results obtained by exact numerical simulations and within the RWA approach. Our analysis is based on the linear expansion around the stable points and it works well in short-time dynamics. The deviations grow gradually as the particle leaves further from the stable points, 

Combining with Eq.~(\ref{Au}), we have in the long-time limit $t\to\infty$,
$
\overline{ |{\bf Z}_i-{\bf \tilde{Z}}^0_i|^2}=2T.
$
 Therefore, the long-time distribution is a normal distribution with the width $\sigma=\sqrt{2T}$. In order to keep phase space crystal stable, we need the dispersion width much smaller than the characteristic length of unit cell (here we take $\pi/2$)
\begin{eqnarray}\label{eq-sigmaT}
\sigma=\sqrt{2T}\ll \pi/2 \ \ \ \Rightarrow \ \ \  T\ll\pi^2/8.
\end{eqnarray}
In fact, the above relation can be directly obtained when we apply the equipartition theorem to the system in the laboratory frame.  Considering a particle trapped at the bottom of harmonic well subjected to a both with temperature $T$, the width of the thermal ground state is
$\overline{\frac{1}{2} { X}_i^2}+\overline{\frac{1}{2}{ P}_i^2}=T$. Note that we have set the Boltzmann constant $k_B=1$ here.
 Actually, for any fixed point, in the regime where the temperature is so low that only slow motions are thermalized, we have $$\overline{\frac{1}{2}( { X}_i-{ \overline{X}}^0_i)^2}+\overline{\frac{1}{2}( { P}_i-{ \overline{P}}^0_i)^2}=T,$$ and thus $\overline{ |{\bf Z}_i-{\bf \overline{Z}}^0_i|^2}=2T$,
where ${\bf \overline{Z}}^0_i$ is the average phase space position of a harmonic oscillator.
 The condition Eq.~(\ref{eq-sigmaT}) just means that the phase space lattice constant (characteristic length of driving lattice) has to be much larger than the width of thermal state.

%

\begin{figure*}
\center
\includegraphics[scale=0.9]{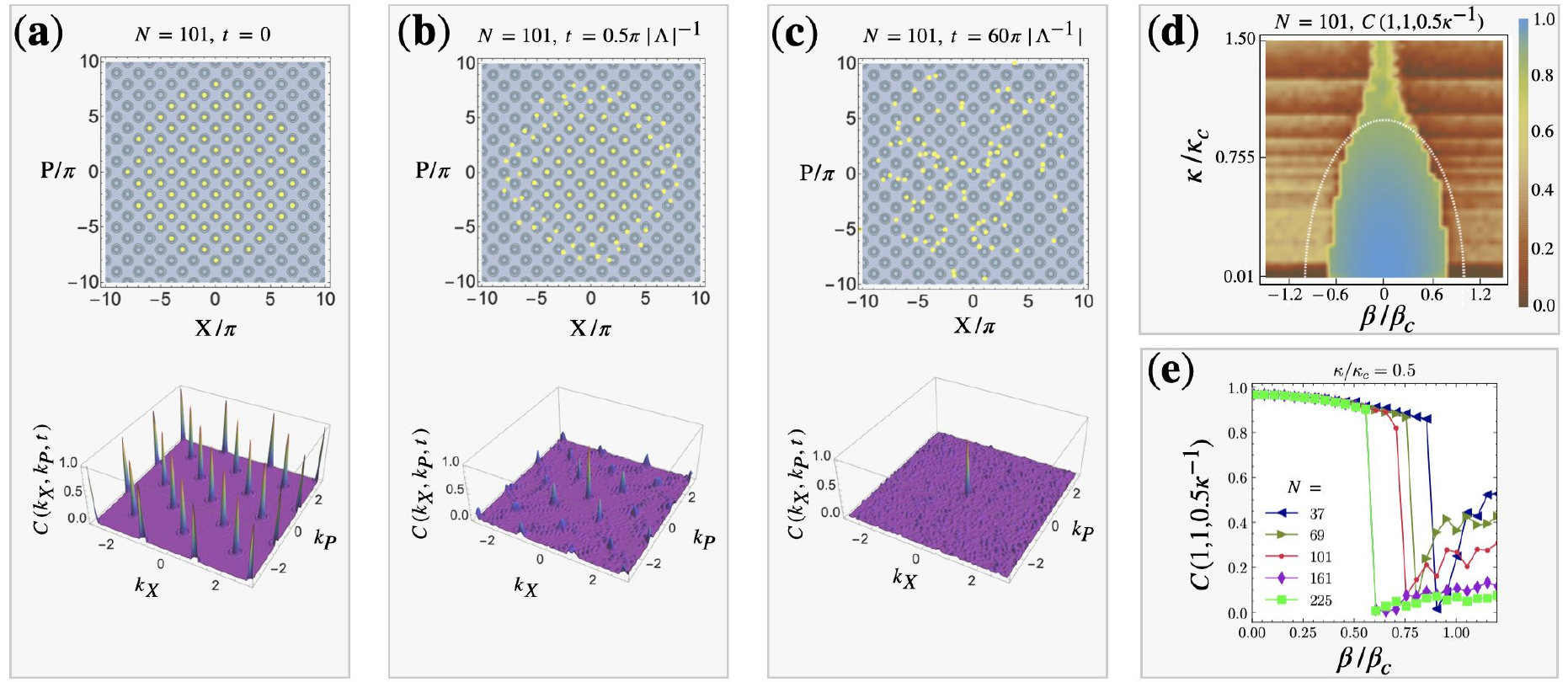}
\caption{{\bf Crystal order parameter and phase diagram.}
 {\bf(a)}-{\bf(c):} Configuration of $N=101$ atoms in phase space (upper {panels}) and the crystal {order} parameter $C(k_X,k_P,t)$ in $(k_X,k_P)$-space (lower {panels}) at time instant $t=0$ (a), $t=0.5\pi|\Lambda^{-1}|$ (b) and $t=60\pi|\Lambda^{-1}|$ (c) with $\Lambda=-0.01$,  interaction strength $\beta=1.5\beta_c$, dissipation rate $\kappa=0$ and temperature $T=0$. {\bf (d):}  Phase diagram, i.e. the crystal order parameter  $C(k_X=1,k_P=1,t=\frac{1}{2\kappa})$ as a function of the scaled interaction strength $\beta/\beta_c$ and the scaled dissipation rate $\kappa/\kappa_c$. The atom number is $N=101$ and the temperature is zero ($T=0$). {The yellow curve corresponds to the analytical prediction (\ref{interaction_dissipation_condition}).}
 {\bf (e)} Crystal order parameter $C(1,1,\frac{1}{2\kappa})$ as a function of the interaction strength for different atom numbers $N=37$, 69, 101, 161, 225. The dissipation rate is $\kappa=0.5\kappa_c$. Note that the curves for $N=161$ and $225$ are nearly overlaping each other. 
 }\label{figPTbetakappa}
\end{figure*}

\subsection{Interaction effects}
We now study the interaction effects on the existence of phase space crystals. For convenience, we separate the total Hamiltonian into the sum of two parts
\begin{eqnarray}\label{}
{\cal H}&=&\sum_i\Lambda(\cos X_i+\cos P_i)^2+\sum_{i<j}U(R_{ij})\nl
&\equiv& \mathcal{T}+\Phi.
\end{eqnarray}
Here, we have introduced $\mathcal{T}\equiv\sum_i\Lambda(\cos X_i+\cos P_i)^2$ representing the summary of all single-particle square lattice Hamiltonians, and $\Phi\equiv\sum_{i<j}U(R_{ij})$ representing the sum of the interaction potentials. In the presence of the dissipation but at zero temperature $T=0$, the fixed points are given by the condition $ {\bf F}_i=0$, cf. Eq.~(\ref{eq-FikT}), 
\begin{eqnarray}\label{}
{\bf \nabla}_i\mathcal{T} \big|_{{\bf Z}_i={\bf \tilde{Z}}^0_i}=- {\bf \nabla}_i\Phi \big|_{{\bf Z}_i={\bf \tilde{Z}}^0_i}+\frac{1}{2}\kappa \big({\bf \hat{n}}\times{\bf \tilde{Z}}^0_i\big),
\end{eqnarray}
where $ {\bf \tilde{Z}}^0_i$ is the shifted fixed points to be calculated. By linearizing the above condition around the original fixed points ${\bf Z}^0_i$ satisfying ${\bf \nabla}_i\mathcal{T} |_{ {\bf Z}_i={\bf Z}^0_i}=0$, we obtain
\begin{eqnarray}\label{}
{\bf J}_\mathcal{T}( {\bf \tilde{Z}}^0_i-{\bf Z}^0_i)&=&- {\bf \nabla}_i\Phi({\bf Z}^0_i)-{\bf J}_{\Phi}( {\bf \tilde{Z}}^0_i-{\bf Z}^0_i)\nl
&&+\frac{1}{2}\kappa {\bf A} ({\bf \tilde{Z}}^0_i-{\bf Z}^0_i)+\frac{1}{2}\kappa {\bf A} {\bf Z}^0_i,
\end{eqnarray}
where ${\bf J}_\mathcal{T}$ and ${\bf J}_{\Phi}$ are Jacobian tensors defined by
\begin{eqnarray}
&&{\bf J}_\mathcal{T}\equiv{\bf J}_{\nabla_i\mathcal{T}}|_{ {\bf Z}_i={\bf Z}^0_i}\equiv\left(\begin{array}{cc}\frac{\partial^2 \mathcal{T}}{\partial X_i\partial X_i} & \frac{\partial^2\mathcal{T}}{\partial X_i\partial P_i} 
\\\frac{\partial^2\mathcal{T}}{\partial P_i\partial X_i} & \frac{\partial^2 \mathcal{T}}{\partial P_i\partial P_i}\end{array}\right)\Bigg\vert_{ {\bf Z}_i={\bf Z}^0_i}, \ \ \ \ \ \ \ 
\\  
&&{\bf J}_\Phi\equiv{\bf J}_{\nabla_i\Phi}|_{ {\bf Z}_i={\bf Z}^0_i}\equiv\left(\begin{array}{cc}\frac{\partial^2 \Phi}{\partial X_i\partial X_i} & \frac{\partial^2\Phi}{\partial X_i\partial P_i} 
\\\frac{\partial^2\Phi}{\partial P_i\partial X_i} & \frac{\partial^2 \Phi}{\partial P_i\partial P_i}\end{array}\right)\Bigg\vert_{ {\bf Z}_i={\bf Z}^0_i}.
\end{eqnarray}
Therefore, we have the shifted equilibrium points 
\begin{eqnarray}\label{eq-ZiZ0JT}
{\bf \tilde{Z}}^0_i-{\bf Z}^0_i=\big({\bf J}_\mathcal{T}+{\bf J}_\Phi-\frac{1}{2}\kappa {\bf A}\big)^{-1}\big[\frac{1}{2}\kappa {\bf A} {\bf Z}^0_i- {\bf \nabla}_i\Phi({\bf Z}^0_i)\big].\nl
\end{eqnarray}
It is straightforward to calculate tensor ${\bf J}_\mathcal{T}$, cf., Eq.~(\ref{eq-JDeta}). The difficulty is to calculate ${\bf J}_\Phi$ and ${\bf \nabla}_i\Phi({\bf Z}^0_i)$. Below, we provide an approximate method to calculate them analytically.

For the Coulomb-like phase space interaction potential $U(R_{ij})=\pi^{-1} \beta/R_{ij}$, cf., Eq.~(\ref{eq-URij}), the parameter $\beta$ plays the role of an effective charge.  We assume that $N$ atoms are initially uniformly distributed in a disk shape with radius $R$ with density $\rho\equiv N/(\pi R^2)$. We smear the point charges into a uniform charge distribution with charge density $\beta\rho$. Then, the interaction potential at the edge of the disk is given by
\bea
\Phi(R)&=&\int_{-\pi/2}^{\pi/2}d\theta\int_0^{2R\cos\theta}rdr\frac{\beta\rho}{\pi r}
\nl
&=&\frac{1}{\pi}\beta\rho\int_{-\pi/2}^{\pi/2} d\theta \int_0^{2R\cos\theta}dr\nl
&=&\frac{4}{\pi}\beta\rho R.
\eea
Thus, the gradient of interacting potential at the edge is 
\bea\label{eq-edgeforce}
 {\bf \nabla}_i\Phi(R)=\frac{4}{\pi}\beta\rho {\bf \hat{n}^0_i}\ \ \ \mathrm{and} \ \ \ \ {\bf J}_\Phi=0,
\eea
where $ {\bf \hat{n}^0_i}$ is the unit direction from the center of the disk to initial equilibrium position of $i$-th atom, i.e., ${\bf Z}^0_i=Z^0_i{\bf {\hat n}}^0_i$. Note that the phase space force  given by Eq.~(\ref{eq-edgeforce}) is only for the atoms at the edge and independent of the radius $R$.

Using Eqs.~(\ref{JandA})-(\ref{ZiZ02}) and (\ref{eq-ZiZ0JT}), we have the analytical expression for the shifted fixed point as follows
\begin{eqnarray}\label{}
{\bf \tilde{Z}}^0_i-{\bf Z}^0_i
&=&- \frac{\kappa(4\Lambda) Z^0_i/2+2\kappa\pi^{-1}\beta\rho}{(4\Lambda)^2+\kappa^2/4}{\bf \hat{n}} \times{\bf {\hat n}}^0_i\nl
&&-\frac{\kappa^2Z^0_i/4-4(4\Lambda)\pi^{-1}\beta\rho}{(4\Lambda)^2+\kappa^2/4}{\bf {\hat n}}^0_i.
\end{eqnarray}
Comparing to Eq.~(\ref{ZiZ02}), the interaction basically gives a correction to the effect of dissipation. The new equilibrium position is given by
\bea\label{newEP}
{\bf \tilde{Z}}^0_i&=&- \frac{\kappa(4\Lambda) Z^0_i/2+2\kappa\pi^{-1}\beta\rho}{(4\Lambda)^2+\kappa^2/4}{\bf \hat{n}} \times{\bf {\hat n}}^0_i\nl
&&+\frac{(4\Lambda)^2Z^0_i+4(4\Lambda)\pi^{-1}\beta\rho}{(4\Lambda)^2+\kappa^2/4}{\bf {\hat n}}^0_i.
\eea
Following Eq.~(\ref{exiscond1}), we have the existence condition of phase space crystals with interaction
\begin{eqnarray}\label{exiscond2}
&&\Big|- 4\pi^{-1}\beta\rho {\bf \hat{n}^0_i}+\frac{1}{2}\kappa \big({\bf \hat{n}}\times{\bf \tilde{Z}}^0_i\big)\Big|=|{\bf \nabla}_i\mathcal{T}|\nl
&&=2 \Big|\Lambda(\cos X_i+\cos P_i)\Big|\sqrt{\sin^2 X_i+\sin^2 P_i}\nl
&&\leqslant 2\sqrt{2}|\Lambda|.
\end{eqnarray}
As in Eq.~(\ref{exiscond1}), by modifying the upper limit by $2|\Lambda|$ and plugging Eq.~(\ref{newEP}) to Eq.~(\ref{exiscond2}), we have the existence condition for stable phase space crystal state
\begin{eqnarray}\label{exiscond3}
\big(\frac{4\beta\rho}{\pi}\big)^2+\big(\frac{\kappa}{2}\big)^2\Bigg[\frac{(4\Lambda)^2Z^0_i+4(4\Lambda)\frac{\beta\rho}{\pi}}{(4\Lambda)^2+\big(\frac{\kappa}{2}\big)^2}\Bigg]^2\leqslant (2\Lambda)^2.\ \ 
\end{eqnarray}
By introducing the critical interaction strength and the critical dissipation rate as follows 
\bea\label{eq-betac-kappac}
\beta_c\equiv\frac{\pi|\Lambda|}{2\sigma_a}=\pi^3|\Lambda|,\ \ \  
\kappa_c\equiv\frac{4|\Lambda|}{Z^0_i}=\frac{4|\Lambda|}{\sqrt{2\pi N}},
\eea
where we have assumed that one atom is present per each fixed point, i.e. $\rho=\sigma_a=2/(2\pi)^2$ and $\pi (Z^0_i)^2\sigma_a=N$, the existence condition (\ref{exiscond3}) becomes
\begin{eqnarray}\label{exiscond4}
&&\big(\frac{\beta}{\beta_c}\big)^2+\big(\frac{\kappa}{\kappa_c}\big)^2
\Bigg[\frac{1+\frac{\beta}{\beta_c}\frac{1}{2Z^0_i}}{1+\big(\frac{\kappa}{\kappa_c}\big)^2\big(\frac{1}{2Z^0_i}\big)^2}\Bigg]^2\leqslant 1.
\end{eqnarray}
For large lattice $|Z^0_i|\gg1$, we have a simple condition 
\bea
\label{interaction_dissipation_condition}
\big(\frac{\beta}{\beta_c}\big)^2+\big(\frac{\kappa}{\kappa_c}\big)^2\lesssim 1.
\eea
Although this condition for the existence of phase space crystals is obtained based on the linear analysis of dynamical system and other approximations, it provides a very good estimation for the phase transition as shown below by our numerical simulations. When the condition (\ref{interaction_dissipation_condition}) breaks down, the atoms at the edge first escape their stable points and the entire crystal starts to melt from its edge, cf. Fig.~\ref{figPTbetakappa}(b).

\begin{figure}
\center
\includegraphics[scale=0.4]{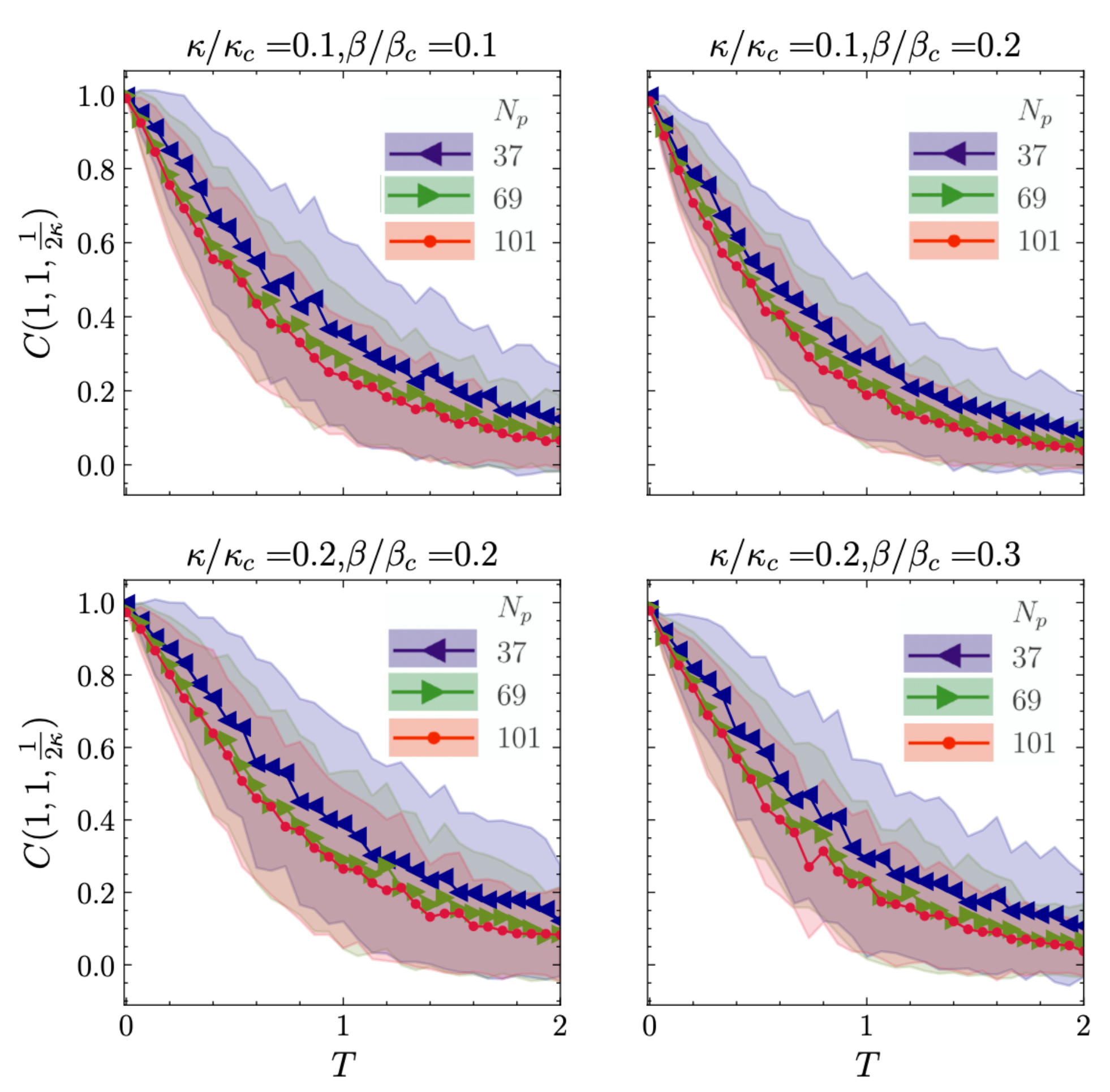}
\caption{{\bf  Crystal order parameter as functions of temperature} for different atom number with given dissipation rate and interaction strength. Connected points are the averaged crystal order parameter $C(1,1,\frac{1}{2\kappa})$ over $N_{iter}=200$ trajectories with coloured shadows indicating the standard deviation $\sigma_C$. The standard error from the mean over $N_{iter}=200$ trajectories is $\sigma_C/\sqrt{N_{iter}}$. }\label{fig-Temperature}
\end{figure}

\subsection{Phase diagram}

To identify the existence of phase space crystal, we define the crystal order parameter as follows
\bea
C(k_X,k_P,t)\equiv\Big|\frac{1}{N}\sum_je^{ik_XX_j(t)+ik_PP_j(t)}\Big|^2.
\eea
In Fig.~\ref{figPTbetakappa}, we plot the crystal order parameter for different system parameters. 

{Let us first illustrate what happens with the crystal when there is no dissipation ($\kappa=0$) but the interaction strength is larger than the critical value.}
In Fig.~\ref{figPTbetakappa}(a), we start from an initial state where $N=101$ atoms occupy the lattice sites in a finite disk-shape region. The corresponding crystal order parameter as a function of $k_X$ and $k_P$ is also shown in Fig.~\ref{figPTbetakappa}(a). For such perfect crystal state, the order parameter plot contains regular peaks that are periodically arranged in the $(k_X,k_P)$-space, i.e., the positions of peaks appear at points $(n,m)$ with $n,m\in\mathbb{Z}$ having the same parity.  In Fig.~\ref{figPTbetakappa}(b), we plot the configuration of $N=101$ atoms in phase space and the crystal parameter in  $(k_X,k_P)$-space at time instant $t=0.5\pi|\Lambda^{-1}|$ with $\Lambda=-0.01$. It is clearly shown that {for the interaction strength $\beta/\beta_c=1.5$ and dissipation rate $\kappa=0$ we consider here,} the crystal starts to melt from the edge as predicted above, cf. Eq.~(\ref{interaction_dissipation_condition}) and the related discussion around. The corresponding crystal order parameter plot in the lower panel of Fig.~\ref{figPTbetakappa}(b) shows all the peaks diminish except the trivial peak at the center $(k_X=0,k_P=0)$. In Fig.~\ref{figPTbetakappa}(c), we plot the configuration of atoms in phase space  and the crystal parameter at time instant $t=60\pi|\Lambda^{-1}|$ with $\Lambda=-0.01$. All the atoms escape from their equilibrium points and spread over the phase space forming a gas-like state in phase space. Because all the atoms are randomly distributed in phase space, all the nontrivial peaks in the crystal order parameter plot disappear. 

{Now, let us include the dissipation but at zero temperature.}
We choose the peak value of crystal order parameter at point $(k_X=1,k_P=1)$ to  
trace the phase diagram. Due to finite dissipation, the atoms need some time to relax to the final state. The characteristic relaxation time scale  is of the order of $1/(2\kappa)$.
In Fig.~\ref{figPTbetakappa}(d), we plot the crystal order parameter  $C(1,1,\frac{1}{2\kappa})$ for $N=101$ atoms as a function of the scaled interaction strength $\beta/\beta_c$ and the scaled dissipation rate $\kappa/\kappa_c$. 
 From the plots, it is clearly  visible that there exists a region in the parameter space spanned by dissipation and interaction {where the order parameter does not vanish}.
In Fig.~\ref{figPTbetakappa}(e), we plot the crystal order parameter as a function of the interaction strength for the dissipation rate $\kappa=0.5\kappa_c$ and for five different atom numbers, i.e. $N=37, 69, 101, 161, 225$. The sudden jump of the order parameter indicates a discontinuous phase transition. As the atom number increases with uniform density (approaching the scenario similar to thermodynamic limit in equilibrium state), the transition point approaches a fixed point close to ({actually a bit} lower than) our predicted value $\beta_c$ based on linear analysis, cf. the transition curves for $N=161$ and $N=225$.  

In order to show the effects of thermal noise on the formation of phase space crystals, we plot {in Fig.~\ref{fig-Temperature}} the crystal order parameter  $C(1,1,\frac{1}{2\kappa})$ as a function of temperature for different dissipation rates and interaction strengths. We show the averaged value (connected points) and the standard deviation (coloured shadows) of the crystal order parameter obtained by simulating 200 different realizations of noise. At low temperature ($T\ll 1$), the order parameter is close to one indicating the existence of a crystal state in phase space. In contrast, at sufficiently high temperature, the order parameter is very close to zero indicating that the crystal state in phase space is totally dissolved. In each plot, we calculate the crystal order parameter for three different system sizes (atom numbers). As atom number increases, the plots approach a fixed curve corresponding to the thermodynamic limit. The standard deviation for each parameter set is zero at zero temperature ($T=0$), and then starts to increase when the temperature increases. But as the temperature becomes high enough, the standard deviation goes back to zero again. This is because the phase space crystal state does not exist and the crystal order parameter {vanishes for all realizations of noise we have simulated.}

\section{Experimental Parameters}

We now discuss whether the conditions for realizing classical phase space crystals can be satisfied in the real cold-atom experiments. We first examine the RWA condition that the driving strength $\Lambda$, dissipation rate $\kappa$ and interaction strength $\beta$ should be much smaller than unity in our units \cite{Guo2016pra,Liang2018njp}. By recovering the units of parameters, we have the RWA condition
\bea\label{eq-rwa-condition}
|\Lambda|,\ \frac{\beta}{l_0}\ll \epsilon_0=m\omega_0^2\Big(\frac{l_0}{2\pi}\Big)^2,\ \ \ \kappa\ll \omega_0.
\eea
In the  experiment of  cold atoms \cite{Moritz2003prl}, a quasi-1D harmonic potential with strong  transverse trapping frequency  is formed by propagating Gaussian laser beam(s). The resulting transverse trapping frequency $\omega_{tr}$ and axial trapping frequency $\omega_{0}$ are given by 
$
\omega_{tr}=\frac{2E_r}{\hbar}\sqrt{\frac{V_0}{E_r}}$, $\omega_{0}={\lambda_L\omega_{tr}}/{\pi w_0},
$
where $w_0$ is the Gaussian beam waist, $\lambda_L$ ($k_L=2\pi/\lambda_L$) is the laser wavelength (wavenumber),  $V_0$ is the intensity of lasers and $E_r=\hbar^2k^2_L/2m$ is the recoil energy of an atom.

For the cold $^{87}Rb$ atoms ($m=\SI{1.42e-25}{\kilogram}$) in the presence of the Gaussian laser with  wavelength $\lambda_L=\SI{823}{\nano\meter}$, beam waist $w_0=\SI{160}{\micro\meter}$ and intensity $V_0=27E_r$,  the transversal and axial trapping frequency are $\omega_{tr}/2\pi= \SI{36}{\kilo\hertz}$ and $\omega_{0}/2\pi= \SI{59}{\hertz}$ respectively. By choosing the characteristic length of driving lattice potential  (created by two additional laser beams, see Fig.~\ref{Fig-Modelsystem}) fifty times of the axial trapping length $l_0=50\sqrt{\hbar/m\omega_0}=\SI{71}{\micro\meter}$, we have the following RWA condition for driving strength 
\bea
|\Lambda|\ll\epsilon_0=1.08E_r
\eea
 with the recoil energy $E_r=2.29\times 10^{-30}\SI{}{\joule}$. Therefore, we can tune the intensity of lasers that generate driving lattice potential  to satisfy the RWA.

In the quasi-1D trap, the effective contact interaction is given by \cite{Bloch2008rmp}
$
V_{1D}(x)\approx 2\hbar\omega_{tr}a_0\delta(x),
$
 where $a_0$ is the 3D $s$-wave scattering length. Thus, we have the following RWA condition for interaction strength from Eq.~(\ref{eq-rwa-condition})
  \bea
\frac{ \beta}{l_0}=\frac{ 2\hbar\omega_{tr}a_0}{l_0}\ll \epsilon_0.
 \eea
 Taking the 3D scattering length $a_0=\SI{5.3}{\nano\meter}$ for  $^{87}Rb$ atoms, we have the interaction strength
$
 \beta\approx 0.00144\epsilon_0 l_0
$
satisfying RWA. The interaction strength can be further tuned either by transversal trapping frequency $\omega_{tr}$ or by the Feshbach resonance \cite{Bloch2008rmp}. 

Next, we estimate the critical parameters for the phase diagram in Fig.~\ref{figPTbetakappa}.  According to Eq.~(\ref{eq-betac-kappac}), we have the critical driving strength and critical dissipation rate with recovered units
\bea
\frac{\beta_c}{\epsilon_0 l_0}=\frac{\pi^3|\Lambda|}{\epsilon_0},\ \ \  \kappa_c=\frac{4|\Lambda|}{\sqrt{2\pi N}}\frac{\omega_0}{\epsilon_0}
\eea
Using the driving strength $|\Lambda|=0.01\epsilon_0$ and atom number $N=225$, we have the critical values $ \beta_c\approx 0.31 \epsilon_0 l_0$ and $\kappa_c\approx 0.0038\omega_0$. The dissipation rate (damping coefficient) $\kappa/\omega_0$ for an atom can be tuned by the laser detuning from atomic frequency and set $\kappa=0$ at resonance \cite{metcalf2007book}.
 Finally, to have stable phase space crystals, we need the temperature condition from Eq.~(\ref{eq-sigmaT})
\begin{eqnarray}\label{}
  T\ll\frac{\pi^2}{8}\frac{ \epsilon_0}{k_B}\approx \SI{179}{\nano\kelvin},
  \end{eqnarray}
which locates in the typical temperature range from the nanokelvin to the microkelvin regime in the cold-atom experiments \cite{Bloch2008rmp}.


\section{Summary and Discussion}\label{sec-Summary}

Many-body phase space crystal is an ordered highly excited state in a classical or quantum many-body dynamical system. Previous works on phase space crystals are restricted to closed system. In this work, we investigated the  dynamics of classical many-body phase space crystals in the open dissipative environment with thermal noise. We started from the exact equations of motion of the system in the lab frame in the presence of dissipation and {at non-zero} temperature. We then derived and justified the equations of motion obtained within the rotating wave approximation in the rotating fame, which describes the slow dynamics of harmonic oscillation's quadratures. 
We performed linear analysis of stability of the phase crystal and found that strong dissipation, interaction and high temperature can destroy the crystal state in phase space. We estimated the critical values of the parameters for the destruction of the phase space crystal.
By defining a crystal order parameter, we plotted the phase diagram in the dissipation-interaction parameter plane and the order parameter as a function of temperature. The main conclusion is that phase space crystal state does exist for a range of parameter settings in the cold atom experiments.

In order to prepare such phase space crystals, one can initially set the cloud of atoms with driving, dissipation and interaction parameters below the critical values according to our prediction, but at relatively high temperature. In this scenario, the thermal noise will activate the atoms spreading over the phase space. Then, when cooling down the atoms, the finite dissipation will help the atoms to relax to the stable points nearby forming some blocks of phase space crystals. As the main goal of the present work is to prove the existence of phase space crystal state, we will study in detail how to prepare phase space crystals in the future work.

In this work, we have studied the square phase space crystalline structure for the single-particle Hamiltonian. An extension to other lattice structure like honeycomb lattice is straightforward. It has been shown that the phase space crystal vibrational band structure of honeycomb lattice can support chiral transport without breaking time-reversal symmetry \cite{Guo2022prb}. Such kind of anomalous Chern insulator has not yet been realised in the experiments. We only investigated dynamics of phase space crystals in classical regime. The extended study in the quantum regime, which is closely related to (fractional) quantum Hall physics and 1D anyons \cite{Tosta2021pra,Greschner2015prl}, will be our future work.

\begin{acknowledgments}
 We acknowledge helpful discussions with Vittorio Peano and Florian Marquardt. {Support of the National Science Centre, Poland, via Project
No. 2018/31/B/ST2/00349 (A.E.K.) is acknowledged. This research was also funded in part by the National Science Centre, Poland, Project No. 2021/42/A/ST2/00017
(K.S.). For the purpose of Open Access, the author has
applied a CC-BY public copyright licence to any Author
Accepted Manuscript (AAM) version arising from this
submission. Numerical computations in this work were supported in
part by PL-Grid Infrastructure.}
\end{acknowledgments}

\appendix

\section{Square phase space lattice}

To show how to generate the single-particle square phase space lattice in Eq.~(\ref{eq-square}), we start from the following generalised model of a kicked harmonic oscillator
\bea\label{A1}
H&=& \frac{1}{2}(x^2+p^2)\nl 
&&+ \sum_{n\in \mathbb{Z}}\sum_{q}K_q \cos(k_qx-\phi_q) \delta(\frac{t}{\tau}-\theta_q -n),\ \ \ 
\eea
where $q$ represents the kicking sequence of stroboscopic lattice with tunable intensity $K_q$, wave vector $k_q$ and phase $\phi_q$ at different time instance $t=\tau(n+\theta_q)$ with $n\in \mathbb{Z}$. To simplify the discussion, we first consider a single kicking sequence, i.e.,
\bea\label{A2}
H_s&=& \frac{1}{2}(x^2+p^2) \nl
&&+ \sum_{n\in \mathbb{Z}}K_q \cos(k_qx-\phi_q) 
\delta(\frac{t}{\tau}-\theta_q -n).
\eea 
We transfer the above Hamiltonian into a rotating frame with the kicking frequency $2\pi/\tau$ using the generating function of the second kind
\begin{equation}\label{A3}
G_2(x,P,t)=\frac{xP}{\cos(2\pi t/\tau)}-\frac{x^2}{2}\tan\Big(\frac{2\pi}{\tau}t\Big)-\frac{P^2}{2}\tan\Big(\frac{2\pi}{\tau}t\Big),\nn
\end{equation}
which results in the transformation of phase space coordinates,
\bea\label{A5}
x &=& P \sin(\frac{2\pi t}{\tau}) + X \cos(\frac{2\pi t}{\tau}),\cr
	 p &=& P \cos(\frac{2\pi t}{\tau}) - X \sin(\frac{2\pi t}{\tau}).
\eea
and the transformed Hamiltonian
\bea\label{A6}
H_s(X,P,t)&=&\frac{1}{2}\delta\omega(X^2+P^2)+\sum_{n\in \mathbb{Z}}K_q \delta(\frac{t}{\tau}-\theta_q -n)\cr
&&\cos(k_q[P \sin(\frac{2\pi t}{\tau}) + X \cos(\frac{2\pi t}{\tau})]-\phi_q). \cr &&
\eea
Here, $\delta\omega\equiv1-2\pi/\tau$ is the detuning between the kicking and harmonic oscillator frequencies. For weak resonant driving ($|K_q|\ll1, \tau=2\pi$), the single-particle dynamics can be simplified by averaging the Hamiltonian over the fast harmonic oscillations. The effective slow dynamics of quadratures is given by the lowest
order Magnus expansion, i.e., the time average of $ H_s(X,P,t)$ over one kicking period,
\begin{eqnarray}\label{}
\mathcal{H}_s(X,P)&=&\frac{1}{\tau}\int_0^{\tau} H_s(X,P,t) dt  \nl
&=&K_q \cos(k_q[P \sin(2\pi\theta_q) + X \cos(2\pi\theta_q)]-\phi_q).\nl 
\end{eqnarray}
Including all the kicks in Eq.(\ref{A1}), we obtain the general form of the phase space lattice Hamiltonian
\begin{eqnarray}\label{A7}
\mathcal{H}_s= \sum_q K_q \cos(k_q[P \sin(2\pi\theta_q) + X \cos(2\pi\theta_q)]-\phi_q).\nl
\end{eqnarray}
In principle, any arbitrary lattice Hamiltonian in phase space can be synthesised by multiple stroboscopic lattices. For the square lattice considered in our work, we can get the desired driving parameters by decomposing the square lattice into a series of cosine functions, i.e., 
\begin{eqnarray}\label{A8}
	\mathcal{H}_s(X,P) &=&\Lambda (\cos X+\cos P)^2 \cr
	&=&\Lambda[1 +\frac{1}{2} \cos(2X)+\frac{1}{2} \cos(2P)\cr
    &&+\cos(X+P)+\cos(X-P)]. 
\end{eqnarray}
We decompose each term in the above equation as follows
\begin{eqnarray}
		\frac{1}{2} \cos(2X) &=&\frac{1}{2} \cos(2[X\cos(2\pi) + P\sin(2\pi)]),\cr
		\frac{1}{2} \cos(2P) &=&\frac{1}{2} \cos(2[X\cos(\frac{2\pi}{4}) + P\sin(\frac{2\pi}{4})]), \cr &&
\end{eqnarray}
 and
\begin{eqnarray}
		\cos(X+P)&=&\frac{1}{2}  \cos(\sqrt{2}[X\cos(\frac{2\pi}{8}) + P\sin(\frac{2\pi}{8})])+ \cr && \cr &&
		\frac{1}{2}  \cos(-\sqrt{2}[X\cos(\frac{10\pi}{8}) + P\sin(\frac{10\pi}{8})]),\cr && \cr
		\cos(X-P)&=&\frac{1}{2}  \cos(\sqrt{2}[X\cos(\frac{14\pi}{8}) + P\sin(\frac{14\pi}{8})])+ \cr && \cr &&
		\frac{1}{2}  \cos(-\sqrt{2}[X\cos(\frac{6\pi}{8}) + P\sin(\frac{6\pi}{8})]).  \cr &&
		\label{eq1:3}
\end{eqnarray}
By comparing the above expansion to Eq.~(\ref{A7}), we can generate the square phase space Hamiltonian by choosing the kicking parameters:
\bea
k_q &=&[ \sqrt{2},2,-\sqrt{2},-\sqrt{2},\sqrt{2},2 ]\nl
\theta_q &=& [\frac{1}{8},\frac{2}{8},\frac{3}{8},\frac{5}{8},\frac{7}{8},1],
\eea
and $K_q=\Lambda/2$ and $\phi_q=0$ for all $q$.



%

\end{document}